\documentclass[%
 reprint,
 amsmath,amssymb,
 aps,
 prl,
]{revtex4-2}

\usepackage{graphicx}% Include figure files
\usepackage{dcolumn}% Align table columns on decimal point
\usepackage{bm}% bold math
\usepackage[utf8]{inputenc}
\usepackage[T1]{fontenc}
\begin{document}

\preprint{APS/123-QED}

\title{Electron-phonon interactions in the Andreev Bound States of aluminum nanobridge Josephson junctions}% Force line breaks with \\

\author{James T. Farmer}
\email{jtfarmer@usc.edu}
\author{Azarin Zarassi}
\author{Sadman Shanto}
\author{Darian Hartsell}
\author{Eli M. Levenson-Falk}
\email{elevenso@usc.edu}
\affiliation{Department of Physics, University of Southern California}%
\affiliation{Center for Quantum Information Science and Technology, University of Southern California}

\date{\today}% It is always \today, today,

\begin{abstract}
We report continuous measurements of quasiparticles trapping and clearing from Andreev Bound States in aluminum nanobridge Josephson junctions integrated into a superconducting-qubit-like device. We find that trapping is well modeled by independent spontaneous emission events. Above 80 mK the clearing process is well described by absorption of thermal phonons, but other temperature-independent mechanisms dominate at low temperature. We find complex structure in the dependence of the low-temperature clearing rate on the Andreev Bound State energy. Our results shed light on quasiparticle behavior in qubit-like circuits.
%\begin{description}
%\end{description}
\end{abstract}

\maketitle

%\tableofcontents
Non-equilibrium quasiparticles (QPs) in superconducting quantum circuits can hinder device operation, limiting coherence in most qubit architectures \cite{serniak_hot_2018,vepsalainen_impact_2020} and inducing correlated, difficult-to-correct errors across multiple qubits on the same chip \cite{wilen_correlated_2021,martinis_saving_2021}. The QPs are generated by non-thermal mechanisms such as pair-breaking infrared photons \cite{barends_minimizing_2011} or energy dissipation from local radioactivity and cosmic rays \cite{vepsalainen_impact_2020}. Significant non-equilibrium QP populations with fractional densities $x_{qp}\sim 10^{-9} - 10^{-5}$  are ubiquitously observed \cite{barends_minimizing_2011,de_visser_number_2011,serniak_direct_2019,mannila_superconductor_2022} and have proven difficult to eliminate. %AZ: add reference? There's 4 orders of magnitude range so probably multiple references - Done JF
Mitigation strategies such as improved light-tight shielding \cite{barends_minimizing_2011}, input/output filtering with infrared absorbers \cite{corcoles_protecting_2011, pop_coherent_2014, serniak_direct_2019}, and device engineering \cite{court_quantitative_2008,riwar_normal-metal_2016,henriques_phonon_2019,rafferty_spurious_2021,kurter_quasiparticle_2022,pan_engineering_2022,bargerbos_mitigation_2022} have reduced QP densities over the last decade. 
%However, non-equilibrium QPs are not yet fully understood and remain a problem for the development of superconducting quantum processors. 
Many works have probed QP populations by detecting single charge tunneling across Josephson junctions \cite{aumentado_nonequilibrium_2004,naaman_time-domain_2006,lutchyn_kinetics_2007,shaw_kinetics_2008,sun_measurements_2012,serniak_hot_2018,serniak_direct_2019,kurter_quasiparticle_2022} or observing QPs trapped inside the Andreev Bound States (ABS) of a junction \cite{zgirski_evidence_2011,levenson-falk_single-quasiparticle_2014,olivares_dynamics_2014,hays_direct_2018,farmer_continuous_2021}. These ABS provide a complementary measurement of QP behavior, and can be used as qubit modes themselves \cite{janvier_coherent_2015,hays_coherent_2021}. In many implementations the ABS qubit relies on a non-equilibrium QP trapping in order to initialize the state; such qubits are vulnerable to additional trapping events and to accidental clearing of the QP from the ABS. There is thus a great need to better understand the behavior of QPs in ABS and the mechanisms for QPs transitions between ABS and bulk continuum states.

In this letter we investigate the electron-phonon interactions involved in trapping a quasiparticle into / clearing a quasiparticle from an Andreev Bound State. We show continuous, real-time measurements of ABS trapping dynamics as a function of ABS energy and device temperature in a superconducting-qubit-like device. We find that QP trapping is consistent with independent spontaneous emission events from a bulk QP population that is a combination of a temperature-independent non-equilibrium background and a thermal equilibrium density. We further find that QP clearing from an ABS is consistent with a process dominated by absorption of a thermal phonon at temperatures above 80 mK. At low temperatures, we find evidence that absorption of microwave photons by trapped QPs is the dominant clearing mechanism, even at low drive powers. We analyze the mean QP occupancy of our ABS device and find independent confirmation of our trapping and clearing models. Our results shed light on quasiparticle behavior in ABS and in qubit-like circuits in general.

To study QP trapping, we require a circuit element which is sensitive to the occupation of single electron states with tunability below the superconducting gap. We find such an element in the aluminum nanobridge Josephson junction, an all-superconducting junction which was shown \cite{vijay_optimizing_2009,vijay_approaching_2010,levenson-falk_nonlinear_2011} to follow the KO-1 current-phase relation \cite{kulik_contribution_1975} while providing several hundred conduction channels. Each conduction channel hosts a pair of ABS with energies 
\begin{equation}
    E_A(\delta) = \pm\Delta\sqrt{1-\tau \sin{\frac{\delta}{2}}}
\end{equation}
measured from the Fermi energy.
The transparency $\tau$ is the probability that an incident Cooper pair is transmitted across the junction and $\delta$ is the phase bias across the junction; $\Delta$ is the superconducting gap.
For short ($\lesssim 100$ nm) aluminum nanobridges, $\tau$ approximately follows a Dorokhov distribution with a strong preference to be 0 or 1 \cite{vijay_approaching_2010,dorokhov_transmission_1982}.
When occupied, each ABS in a given channel carries equal and opposite contributions to the supercurrent.
The negative state is usually occupied while the positive state is unoccupied. However, the positive ABS dips below the gap $\Delta$ when both $\delta$ and $\tau$ are nonzero, making it energetically favorable for a quasiparticle above the gap (i.e.~in the bulk continuum) to relax into the ABS and become trapped.
When this occurs the supercurrent contribution of the given channel is cancelled and the channel is ``poisoned".
This is the mechanism of our detection: the Josephson inductance becomes a function of the number of trapped quasiparticles.

By embedding a DC SQUID with symmetric aluminum nanobridge junctions in a $\lambda/4$ coplanar waveguide resonator, we are able to measure the trapping of a QP as a resonant frequency shift of the resonator.
This allows for a high bandwidth, continuous measurement of the ABS occupation in a qubit-like circuit using a standard dispersive measurement setup \cite{farmer_continuous_2021}. A constant flux bias on the SQUID introduces a constant, symmetric phase bias to the junctions $\delta = \pi\phi$ (where $\phi$ is the applied flux in units of flux quanta), tuning the ABS energies.
The fundamental mode $f_0(\phi)$ of our resonator is flux tunable from 4.301 GHz to \~4.25 GHz with a linewidth $\kappa = 2\pi \times 250$ kHz and the shift due to trapping a single quasiparticle $\chi(\phi)/2\pi$ ranges from 100 kHz to 400 kHz.

We perform continuous microwave reflection measurements on our device which is mounted in a dilution refrigerator with a base temperature of 30 mK.
The reflected signal is homodyne demodulated with an IQ mixer and the two quadratures of signal are recorded as a gapless voltage record in 3 s segments by an Alazar 9371 digitizer operating at 300 MHz sample rate. This is down sampled to 1 MHz sample rate before fitting each record
%saving.
%Each segment of raw data is fit 
to a Hidden Markov Model (HMM) \cite{noauthor_hmmlearn_nodate,rabiner_tutorial_1989} %AZ: reference not showing in pdf -fixed JF
with further down sampling if required to maintain signal to noise ratio greater than 3. 
The HMM analysis models the tranisition rates between each trapped QP number and is used to extract a time series of the number of trapped QPs from the continuous voltage record. 
Data was collected and processed in this way over a range of parameters: the dilution refrigerator temperature, the ABS energy, and the applied microwave power. 
For brevity, we restrict ourselves in this analysis to a constant power of -133 dBm ($\sim$ 25 photons) at the device. More details are given in the Supplement \cite{supplement}.

We present three quantities of interest:
$\Gamma_{trap}$ is the rate of QPs relaxing from the bulk into available ABS of the junction, 
$\Gamma_{release}$ is the rate of clearing QPs from ABS to the bulk,
and $\bar{n}$ which we call the mean occupation is the time average of the number of trapped QPs.
$\Gamma_{trap}$ and $\Gamma_{release}$ are found from the off-diagonal elements of the HMM transition matrix\textemdash that is, they are parameters of the model used to extract the ABS occupation time series\textemdash while $\bar{n}$ is found from averaging the extracted occupation over the full 3 second record. 

We begin our modeling with the trap rate.
Assuming trapping events are independent of each other and spontaneous emission dominates the QP relaxation into the ABS, each QP in the bulk has a temperature-independent trapping rate.
This implies the overall trap rate is separable: $\Gamma_{trap}(\Delta_A,T) = f(\Delta_A)x(T)$, where $x(T)$ is the fractional quasiparticle density and $\Delta_A\equiv \Delta-E_A$
%\begin{equation}
%\Delta_A \equiv \Delta\left(1 - \sqrt{1-\tau %\sin^2\frac{\delta}{2}}\right)
%\end{equation}
is the trap depth.
%and $\delta = \pi\phi$ is the phase difference across a junction, assuming two symmetric junctions in a SQUID with negligible loop inductance. 
We take the limit $\tau\rightarrow 1$ as the Dorkhov distribution $\rho(\tau)$ is sharply peaked at 0 and 1, and channels with 0 transmittivity do not contribute to the transport.
The fractional quasiparticle density should be the sum of a non-equilibrium background $x_{ne}$ and a thermal population: 
\begin{equation}
x(T) = x_{ne} + \sqrt{\frac{2\pi k_B T}{\Delta}}\exp{\left(\frac{-\Delta}{k_B T}\right)}.
\end{equation}
We expect that most bulk QPs are near the gap energy, so for spontaneous emission we take $f(\Delta_A) \propto \Delta_A^3$. Putting this together, we obtain a model for the trap rate
\begin{equation}
    \Gamma_{trap} = \beta\Delta_A^3\left(x_{ne} + \sqrt{\frac{2\pi k_B T}{\Delta}}\exp\left(\frac{-\Delta}{k_B T}\right)\right)
    \label{eq:traprate}
\end{equation}
where $\beta$, $\Delta$, and $x_{ne}$ are the free parameters.
To improve the quality of our fit, we take advantage of the low temperature saturation of trap rate $\Gamma_{trap}^0(\Delta_A) \approx \beta\Delta_A^3x_{ne}$ for $T \leq 120$ mK.
We first subtract $\Gamma_{trap}^0(\Delta_A)$ from Eq. \ref{eq:traprate} and fit the resulting quantity to find the gap $\Delta$ and scaling factor $\beta$.
Next we divide Eq. \ref{eq:traprate} by $\Gamma_{trap}^0(\Delta_A)$ and fit this normalized rate with the fractional non-equilibrium density $x_{ne}$ as the only free parameter.
This fitting procedure is covered in detail in the supplement \cite{supplement}.
In Figure \ref{fig:TrapRate}, we show the full model (Eq. \ref{eq:traprate}) using the combined results of this fitting procedure.
\begin{figure}
    \centering
    \includegraphics{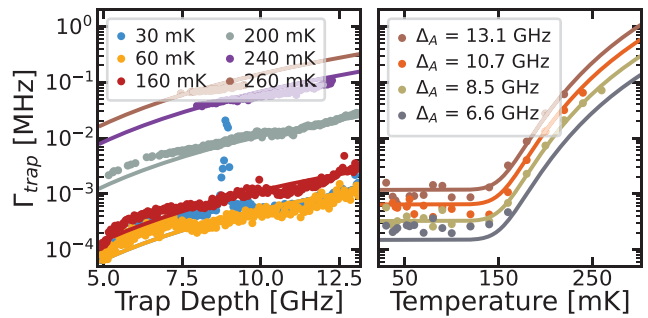}
    \caption{Measured trap rate (circles) and model (solid lines).
    %with fit parameters $\beta = 8.73 \pm 0.68 \times 10^{15}$, fractional non-equilibrium QP density $x_{ne} = 8.5 \pm 0.1 \times 10^{-7}$, and superconducting gap $\Delta = 185 \pm 1.5 \mu$eV. 
    The dependence on the trap depth $\Delta_A$ is shown on the left, while temperature dependence is shown on the right. We note the peak in 30 mK data around 9 GHz on the left was observed as a period of significantly larger than normal mean occupation which lasted approximately 1 hour in laboratory time. The source of this peak has not been found and it is not reproducible.}
    \label{fig:TrapRate}
\end{figure}
We find $\beta = 8.73 \pm 0.68 \times 10^{15}$ MHz/eV$^3$, $x_{ne} = 8.50 \pm 0.10 \times 10^{-7}$, and $\Delta = 185.0 \pm 1.5 \mu\text{eV}$.
We note that the fractional non-equilibrium density $x_{ne}$ is quite high compared to recent works \cite{serniak_direct_2019,vepsalainen_impact_2020,mannila_superconductor_2022} which show a fractional density on the order of $10^{-9}$. 
Our setup uses light-tight radiation shields on all stages of the fridge, with Berkeley Black infrared-absorbing coating \cite{persky_review_1999} on the interior of the 100 mK and mixing chamber shields.
In addition, the sample package is mounted inside of an Amumetal 4K shield with a tin-plated copper can nested inside, also with a Berkeley Black interior coating.
We use custom-made Eccosorb filters as well as K\&L 12 Ghz low-pass filters on all inputs and outputs.
A full diagram is available in Supplementary Figure 1 of the supplement \cite{supplement}. We suspect that our device geometry may contribute to the higher-than-expected density, as large areas of superconducting aluminum are galvanically coupled to the SQUID.

The left panel of Figure \ref{fig:TrapRate} shows a peak in the 30 mK data near 9 GHz.
This anomaly was present in the trap rate and mean occupation, while the release rate was marginally increased. We attribute this to a temporary increase in the bulk QP density, as repeated measurement under nearly identical conditions did not show this effect.
The period of increased trapping lasted for approximately one hour with no change in fridge conditions and no obvious environment factors to blame.
We note the duration of the effect is too long to be caused by adhesive strain \cite{anthony-petersen_stress_2022} or a strong cosmic ray \cite{cardani_reducing_2021}.

We now turn our attention to $\Gamma_{release}$.
To promote a trapped QP from ABS to the continuum of states above the gap, sufficient energy (at least $\Delta_A$) must be absorbed.
In a well shielded dilution refrigerator, we expect this energy to come from the absorption of phonons.
The clearing rate due to electron-phonon interactions should be linear in the phonon density,
\begin{equation}
\Gamma_{phonon}(\Delta_A,T) \propto \rho_{\epsilon \geq \Delta_A}(T),\label{eq:rho}
\end{equation}
where $\rho_{\epsilon \geq \Delta_A}(T)$ is the density of phonons with energy exceeding the trap depth.
In the supplement \cite{supplement}, we integrate the Debye density of states and Bose-Einstien distribution over energies exceeding the trap depth to obtain the model for QP clearing due to phonons:
\begin{widetext}
\begin{equation}
\Gamma_{phonon}(\Delta_A,T) = \alpha T^3 \left[-\left(\frac{\Delta_A}{k_B T}\right)^2 \ln{\left(1-e^{\frac{-\Delta_A}{k_B T}}\right)} + \frac{2\Delta_A}{k_B T} \text{Li}_2\left(e^{\frac{-\Delta_A}{k_B T}}\right) + 2\text{Li}_3\left(e^{\frac{-\Delta_A}{k_B T}}\right)\right]. \label{eq:wideeq}
\end{equation}
\end{widetext}
In the above, $\text{Li}_n (x)$ is the polylogarithm function of order $n$ and $\alpha = C_{ABS\rightarrow bulk}k_B^3/2\pi^2\hbar^3\nu^3$ is an overall scaling factor; $\nu$ is the speed of sound in our sample and $C_{ABS\rightarrow bulk}$ relates the ABS clearing rate to the phonon density. The formal foundation for $C_{ABS\rightarrow bulk}$ is a matter worthy of study as it represents the coupling between ABS and an incoherent bath.

In our measurements, we observe that the release rate saturates at $T\leq$ 60 mK to a value which depends on the power of our microwave readout tone, suggesting that low-temperature clearing is dominated by driven electron-photon interactions. This is surprising because a single readout photon ($\approx$ 4.27 GHz) has insufficient energy to clear the ABS trap ($\Delta_A(\phi) > 5$ GHz $\forall$ measured $\phi$).
Accounting for this readout-dominated electron-photon clearing, we can model the total release rate as
\begin{equation}
\Gamma_{release}(\Delta_A,T) = \Gamma_{RO}(\Delta_A) + \Gamma_{phonon}(\Delta_A,T),
\end{equation}
where the electron-photon clearing rate $\Gamma_{RO}$ is the subject of future work.
For now, we take advantage of the low temperature saturation $\Gamma_{release}^0 \approx \Gamma_{RO}$ to eliminate this photon contribution and maintain focus on the electron-phonon clearing rate.
Our model is 
\begin{equation}
    \Gamma_{release}(\Delta_A,T) - \Gamma_{release}^0(\Delta_A) \approx \Gamma_{phonon}(\Delta_A,T) \label{eq:RelEq}
\end{equation}
which is equivalent to the right hand side of Eq. \ref{eq:wideeq}.
%In practice, the low temperature release rate $\Gamma_{release}^0(\Delta_A)$ is taken as the average over data 60 mK and below.
\begin{figure}[h!]
    \centering
    \includegraphics{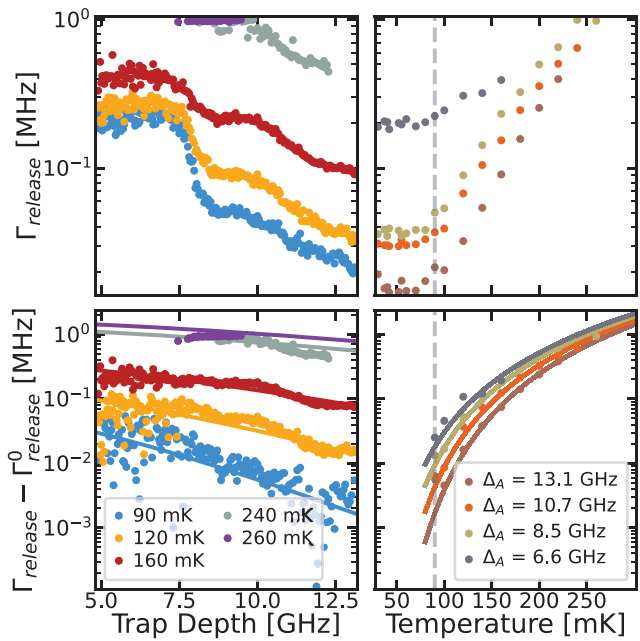}
    \caption{(top) The measured release rate vs trap depth and temperature. The top left panel shows structure in the trap depth dependence which is attributed to the driven electron-photon interactions which dominate at low temperature. In the top right panel, the low temperature saturation is visible. The grey dashed line indicates the cutoff temperature (90 mK) for the fit. (bottom) The measured release rate minus the low temperature saturation is shown as circles, while the phonon clearing model (Eq. \ref{eq:RelEq}) is shown as solid curves.}
    \label{fig:RelRate}
\end{figure}
We keep $\Delta = 185 \mu$eV and fit Eq. \ref{eq:RelEq} with $\alpha$ as the only free parameter as shown in Figure \ref{fig:RelRate}. 
We obtain $\alpha = 38.51 \pm 0.36 \text{ MHz/K}^3$.
Clearly the high temperature release rate is dominated by a thermal distribution of phonons, but this result shows that non-thermal sources may dominate at typical qubit operating temperatures. 
We point out that the 240 mK and 260 mK data in the top left panel show some clipping of the release rate data to the 1 MHz sample rate -- A limitation of our measurement rather than a physical effect.

\begin{figure}
    \centering
    \includegraphics{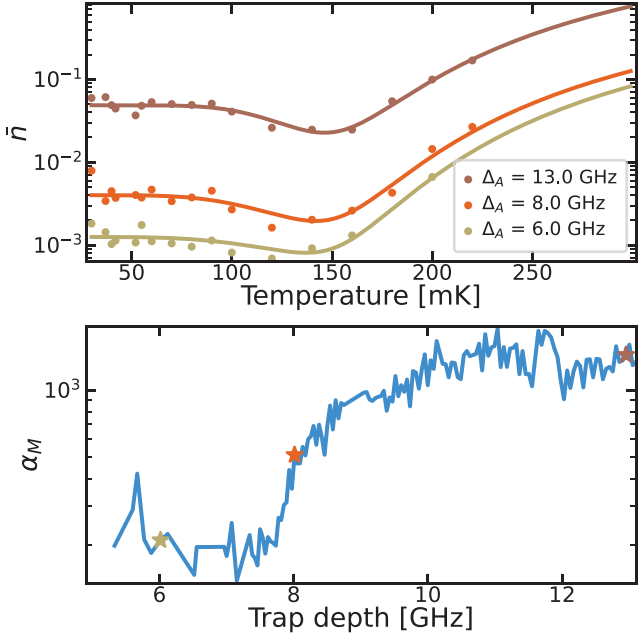}
    \caption{(top) The measured mean occupation (circles) and the corresponding fit (solid) are shown against temperature. Note that a different fit is performed at each value of $\Delta_A$. (bottom) The fit parameter $\alpha_M$ vs trap depth. Stars indicate the value of $\alpha_M$ for the three curves of the same color displayed in the top panel.}
    \label{MeanOcc}
\end{figure}

Our last feature of interest is the mean occupation $\bar{n}$, which is taken directly from the extracted time series of ABS occupations, not from HMM parameters.
We start with a simple sum over weighted probabilities:
\begin{equation}
    \bar{n} = \sum_i i P(i), \label{eq:prob}
\end{equation}
where $P(i)$ is the probability of having $i$ trapped QPs.
In this analysis, we are only distinguishing between 1 trapped QP and 0 trapped QPs, as the incidence of 2 or more trapped QPs is quite rare. We can therefore assume a stationary distribution to obtain
\begin{equation}
    P(0)\Gamma_{trap} = P(1)\Gamma_{release}. \label{eq:stationary}
\end{equation}
Plugging (\ref{eq:stationary}) into (\ref{eq:prob}), we obtain the model for the mean occupation:
\begin{equation}
    \bar{n}(\Delta_A,T) = P(0)\frac{\Gamma_{trap}(\Delta_A,T)}{\Gamma_{release}(\Delta_A,T)}.
\end{equation}
Unfortunately, we are unable to eliminate the driven electron-photon contribution as we did in Eq. (\ref{eq:RelEq}) so we simply leave $\Gamma_{RO}(\Delta_A)$ as a free parameter and fit each line cut along temperature separately. 
We normalize by dividing out the low-temperature saturation ($T\leq 60$ mK) to obtain the model
\begin{widetext}
\begin{equation}
    \Vert\bar{n}_{\Delta_A}(T)\Vert = \frac{1+\frac{1}{x_{ne}}\sqrt{\frac{2\pi k_B T}{\Delta}}e^{\frac{-\Delta}{k_B T}}}{1 + \alpha_M T^3 \left[-\left(\frac{\Delta_A}{k_B T}\right)^2
\ln{\left(1-e^{\frac{-\Delta_A}{k_B T}}\right)} + \frac{2\Delta_A}{k_B T} 
\text{Li}_2\left(e^{\frac{-\Delta_A}{k_B T}}\right) + 2\text{Li}_3\left(e^{\frac{-\Delta_A}{k_B T}}\right)\right]}.
\end{equation}
\end{widetext}
We fit this independently for each trap depth, while holding $x_{ne} = 8.5\times10^{-7}$ and $\Delta = 185 \mu\text{eV}$ fixed. The only fit parameter is $\alpha_M \equiv \alpha / \Gamma_{RO}(\Delta_A)$.
The results are shown in Figure \ref{MeanOcc}.
We note the characteristic dip in mean occupation for $T\in[80,150]$ mK arises from an increased phonon population leading to faster clearing of ABS, while the rise for $T \geq 150$ mK is due to large population of thermal QPs.

\begin{figure}
    \centering
    \includegraphics{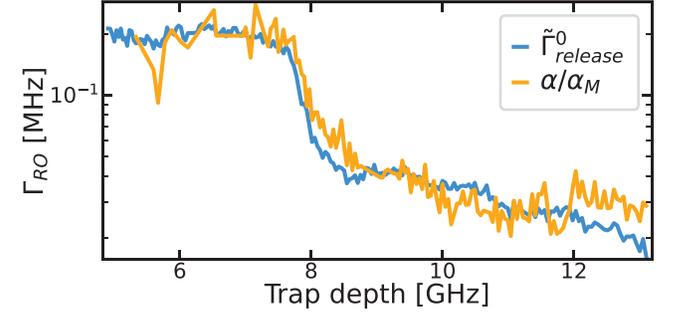}
    \caption{Two sources of estimate for the rate of readout photons clearing QPs from the ABS traps. The measured low temperature release rate (blue) and the fit parameter from the mean occupation, shown as $\alpha/\alpha_M$ (orange), where $\alpha = 38.51$ is found from fitting the phonon contribution to the release rate as shown in Figure \ref{fig:RelRate}. We point out that these agree in shape and magnitude despite coming from different sources.}
    \label{GammaAlpha}
\end{figure}

We may check for self-consistency in our description by examining the relationship between $\alpha_M(\Delta_A)$ and the driven electron-photon clearing rate $\Gamma_{RO}(\Delta_A)$.
We directly measure $\Gamma_{RO}(\Delta_A)$ as the low-temperature saturation of the release rate and compare this with the estimate obtained from $\alpha/\alpha_M$, as shown in Figure \ref{GammaAlpha}. 
Note that the former quantity comes entirely from the HMM parameters, while the latter quantity comes from direct analysis of the ABS occupation time series. These quantities agree very closely, indicating that our analysis is robust.
The driven electron-photon clearing rate has significant structure in its dependence on $\Delta_A$ which is repeatable. There is additional structure when one looks at the dependence on the microwave power, which is the focus of our future work with this system.

By utilizing the many ABS of aluminum nanobridge Josephson junctions, we are able to measure and explain the behavior of quasiparticle trapping in qubit-like circuits over a range of trap depth and temperature.
We show that QPs relax into traps primarily by spontaneous emission of a phonon.
The close agreement between our data and our model suggests that most QPs entering the trap are originally at or near the superconducting gap $\Delta$.
This indicates that any ``hot'' non-equilibrium quasiparticles are first relaxing to the gap in an independent process before trapping or that the majority of non-equilibrium quasiparticles exist at the gap edge, in agreement with past results \cite{diamond_distinguishing_2022}. We do not see any evidence of ``photon-assisted trapping'' (in analogy to the photon-assisted tunneling observed in tunnel junctions) where an infrared photon breaks a Cooper pair, promoting a QP directly into an ABS. This process may occur at lower rates, and is the subject of future work. We also show that clearing of QPs from ABS traps at temperatures above 90 mK occurs primarily through absorption of phonons which are distributed according to the Debye model. Other sources, such as microwave photons, are the dominant source of ABS-clearing energy at qubit operating temperatures. Our results further elucidate the behavior of equilibrium and non-equilibrium quasiparticles in superconducting circuits.

%+++++++++++++++++++++++++++++++++++++++++++++++
%++++.   PAPER END                      ++++++++
%+++++++++++++++++++++++++++++++++++++++++++++++

\begin{acknowledgments}
We thank Leonid Glazman for useful discussions and the MIT Lincoln Laboratory for providing the TWPA used in these measurements.
This work was funded by the AFOSR under FA9550-19-1-0060 and the NSF under DMR-1900135. JF, AZ, DH fabricated the device. JF, AZ, SS performed measurements. JF and AZ performed the analysis. JF wrote the manuscript with input from all authors. ELF was the PI supervising all aspects of the work.
%JF and AZ contributed equally to this work and performed device fabrication, measurement, and data analysis. SS contributed to measurements and DH assisted with fabrication.
%AZ: I don't think JF and AZ contributed equally. I'd be more specific. JF, AZ, DH fabricated the device. JF, AZ, SS performed measurements. JF and AZ ran data analysis. JF wrote the manuscript with input from all authors.

\end{acknowledgments}

% \nocite{*}

\bibliography{quasiparticles_Oct2022}% Produces the bibliography via BibTeX.

\end{document}

\preprint{APS/123-QED}

\title{Supplemental Material: Electron-phonon interactions in the Andreev Bound States of aluminum nanobridge Josephson junctions}% Force line breaks with \\

\author{James T. Farmer}
\email{jtfarmer@usc.edu}
\author{Azarin Zarassi}
\author{Sadman Shanto}
\author{Darian Hartsell}
\author{Eli M. Levenson-Falk}
\email{elevenso@usc.edu}
\affiliation{Department of Physics, University of Southern California}%
\affiliation{Center for Quantum Information Science and Technology, University of Southern California}

\date{\today}% It is always \today, today,

\maketitle

\section{Experiment}
Our $\lambda/4$ resonator is thermally anchored to the base stage of a dilution refrigerator and AC coupled to a microwave feedline which routes incoming and reflected signals through the signal path shown in Supplementary Figure \ref{fig:diagram}.
Our readout tone $\omega_d$ is generated and split at room temperature with half of the signal entering the dilution refrigerator via a series of attenuators and filters and the other half being routed to the LO port of the demodulation mixer.
At the 30 mK stage, the input signal is low-pass filtered by K\&L 12 GHz filters and in-house manufactured Eccosorb filters to reduce pair breaking photons.
The filtered signal is circulated to interact with our device and the reflected signal is circulated towards the output on the right.
Note that the Josephson parametric amplifier (JPA) was not used in this experiment as any attempt to tune it seemed to have a significant effect on the resonance and QP behavior, which we expect was due to insufficient isolation between the JPA and device. For this experiment, the JPA was flux biased to be a few GHz from the resonator frequency.
After some isolation and a low pass filter (Mini Circuits VLF-5850+), the reflected signal is amplified by a travelling wave parametric amplifier (TWPA) whose pump is coupled in after the VLF-5850+ low pass filter.
The amplified reflected signal is then routed through the fridge output, receiving additional amplification by a low noise HEMT at the 4K stage and a chain of isolators and amplifiers at room temperature.
The amplified reflection signal, after coupling -20 dB off to the VNA, is homodyne demodulated with reference to the original signal by the IQ mixer at the top right of panel (b). The DC components of the in-phase and quadrature are low pass filtered and digitized by an AlazarTech ATS 9371 at 300 MHz sample rate.
The raw data is recorded in 3 second segments and downsampled to 5 MHz sample rate prior to saving.
A Keithley 2400 SourceMeter is used to drive current through a coil in the device packaging, coupling magnetic flux through the SQUID loop. As a function of this flux $\phi$, VNA traces are taken to calibrate the drive frequency so as to stay at 1 QP shift below the bare resonance, 
\begin{equation}
    \omega_d(\phi) = 2\pi f_0(\phi) - \chi(\phi).
\end{equation}
\begin{figure*}[h!]
    \centering
    \includegraphics[width=0.95\textwidth]{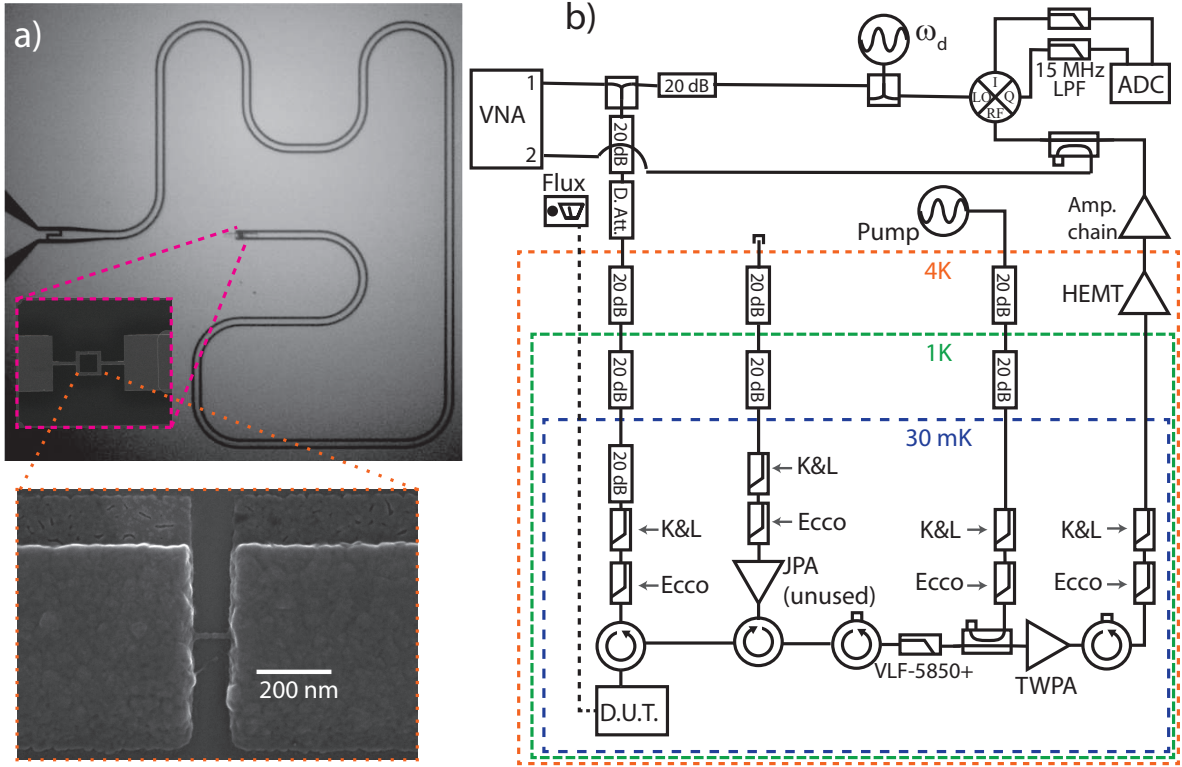}
    \caption{(a) Images of similar device: a $\lambda/4$ resonator is grounded through a DC SQUID with a pair of symmetric aluminum nanowire junctions. These junctions are approximately 25 nm $\times$ 8 nm $\times$ 100 nm. (b) The readout drive $\omega_d$ is generated at room temperature and attenuated along the path through the dilution refrigerator to the base stage. At 30 mK, pair breaking photons on all inputs and outputs are reduced by K\&L 12 GHz low pass filters and Eccosorb CR110 infrared absorbers. The signal circulates to reflect off our device and pass through a 5.85 GHz low pass filter. The signal is then amplified by a travelling wave parametric amplifier (TWPA) whose pump is inserted via a directional coupler. The signal exits the dilution refrigerator receiving further amplification by a HEMT at 4K and a series of low noise amplifiers at room temperature. The signal is homodyne demodulated by an IQ mixer and the resulting quadratures are digitized after 15 MHz low pass filtering. A Keithley Sourcemeter sends DC current along the dashed path to a coil in the device package and a VNA is used to measure the resonance as a function of flux.}
    \label{fig:diagram}
\end{figure*}

Data segments are collected repeatedly while sweeping three parameters: the dilution refrigerator temperature ranges from 30 mK to 300 mK, the flux $\phi \in [0.3,0.5]$ in units of the flux quanta sweeps the ABS trap depth from $\sim5$ GHz to $\sim13$ GHz, and the applied microwave readout power ranges from -155 dBm to -127 dBm at the device.
The temperature is controlled by a PID loop applying current to a resistive heater on the mixing chamber plate and is allowed to stabilize for two hours before starting measurements.
The depth of the ABS, hereafter referred to as the trap depth $\Delta_A$, is modified by an applied magnetic field threading flux through the DC SQUID.
The microwave readout power is generated at 20 dBm and reduced by a series of attenuators from room temperature all the way to 20 mK.
For brevity, we restrict ourselves in this analysis to a constant power of -133 dBm at the device. 
We estimate this readout power to correspond to approximately 25 photons on average based on the single photon readout power of -147 dBm. 

\section{Hidden Markov Model analysis}
Each data record is fit to a Hidden Markov Model (HMM) to obtain the transition matrix $\mathcal{T}$ and Gaussian ``emissions'' $\mathcal{G}$ which best describe the time series of observed data $\mathcal{O} = \{o_0,o_1,o_2,...\}$.
Here we give a brief introduction to HMMs: We assume our system has a set of states $\mathcal{S}$ (corresponding to the different numbers of trapped QPs) and switches from one state to another in a Markovian (memoryless) process with discrete time steps. We cannot directly observe the system state at any step in the time series. What we can observe are the ``emissions'' of our system\textemdash at each point in time, our system emits data (in our case, I and Q voltages) according to a 2D Gaussian probability distribution $G(s)$ which is dependent on the current state $s \in \{\mathcal{S}\}$. 
The system may change between states with probabilities given by the transition matrix $\mathcal{T}$, where the element $T_{ij}$ is the probability of transitioning from state $i$ to state $j$ at the next point in the time series.
Fitting the HMM amounts to choosing the transition matrix $\mathcal{T}$ and emissions $\mathcal{G}$ which maximize the likelihood that our system would produce the sequence of observations $\mathcal{O}$.

Once we have fit the HMM, we can the use the Viterbi algorithm to obtain the sequence of hidden states $\mathcal{N} = \{n_0,n_1,n_2,...\}$ which best matches our data $\mathcal{O}$. In the main text, we refer to this sequence of states $\mathcal{N}$ as the ``extracted occupation'', from which we obtain the mean occupation $\bar{n}$ by averaging $\mathcal{N}$ over the full 3 second record.
The diagonal of $\mathcal{T}$ give the probabilities of staying in the same state, while off-diagonal elements give the probability of transitioning to a new state. 
We can use the off-diagonal elements of $\mathcal{T}$ to get the transition rates by assuming a Poisson process such that the probability of transitioning states is
\begin{equation}
    T_{ij} = \frac{\Gamma_{ij}}{f_s} \exp\left(\frac{-\Gamma_{ij}}{f_s}\right),
\end{equation}
where $f_s$ is the sampling rate.
We can solve for $\Gamma_{ij}$ using the principal branch of the Lambert W function:
\begin{equation}
    \Gamma_{ij} = -f_s \mathrm{W}_0(-T_{ij}).\label{eq:rate}
\end{equation}
We define the trap rate as Eq. \ref{eq:rate} going from $i = 0$ to $j = 1$ trapped QPs,
\begin{equation}
    \Gamma_{trap} \equiv \Gamma_{01}.
\end{equation}
Similarly, the release rate is defined as going from $i =1$ to $j=0$ trapped QPs
\begin{equation}
    \Gamma_{release} \equiv \Gamma_{10}.
\end{equation}

We verified the HMM performance using simulated data and found that the HMM analysis recovers simulated parameters remarkably well so long as we maintain a power SNR (signal-to-noise ratio) greater than 3.
Here we define the SNR as the squared distance between Gaussian centers in the IQ plane (in-phase and quadrature from homodyne demodulation) divided by the product of Gaussian standard deviations. We have also confirmed the stability of our fits with SNR $> 3$ by taking real data with high SNR and adding random Gaussian noise to degrade SNR; all extracted parameters are roughly independent of SNR until it drops below 3.

In processing our data, we start by fitting the HMM to data taken at high measurement power (i.e. high SNR) with $f_s = 1$ MHz sampling rate. We then use this HMM's Gaussian emissions $\mathcal{G}$ as an inital guess to fit data at slightly lower measurement power. We repeat this procedure, bootstrapping our way to lower and lower measurement powers. If at any power the SNR drops below 3, we decrease the sample rate using boxcar downsampling to maintain SNR $\geq 3$. As another check, we ensure that the integration time $1/f_s$ does not exceed half of the shortest mode lifetime $-1/f_s \log(T_{ii})$.
In bootstrapping from high power to low, we continue until we have fit all the data or we are unable to meet the two conditions above. For the lowest temperatures, we are able to bootstrap down to single photon readout or less, while for $T\geq 140$ mK we are unable to reach powers below tens of photons due to the faster transition rates involved.

\section{Fitting the trap rate}
Our model is derived in Eq. 4 of the main text and copied here:
\begin{equation}
    \Gamma_{trap} = \beta\Delta_A^3\left(x_{ne} + \sqrt{\frac{2\pi k_B T}{\Delta}}\exp\left(\frac{-\Delta}{k_B T}\right)\right).
\end{equation}
The thermal contribution is insignificant at low temperatures, so we have 
\begin{equation}
    \Gamma_{trap}^0 \approx \beta\Delta_A^3 x_{ne}.\label{eq:lowTtrap}
\end{equation}
This allows us to calculate two new quantities:
\begin{align*}
    \Gamma_{trap}^- &\equiv \Gamma_{trap} - \Gamma_{trap}^0\\
     &= \beta\Delta_A^3\sqrt{\frac{2\pi k_B T}{\Delta}}\exp\left(\frac{-\Delta}{k_BT}\right)\numberthis\label{eq:subtrap}\\
     \\
    \Vert\Gamma_{trap}\Vert &\equiv \Gamma_{trap}/\Gamma_{trap}^0 \\
     &= 1 + \frac{1}{x_{ne}}\sqrt{\frac{2\pi k_BT}{\Delta}}\exp\left(\frac{-\Delta}{k_BT}\right)\numberthis\label{eq:normtrap}.
\end{align*}
We first fit $\Gamma_{trap}^-$ (Eq. \ref{eq:subtrap}) over the full trap depth and temperature manifold with free parameters $\Delta$ and $\beta$, as shown in Supplementary Figure \ref{fig:subtrap}.
\begin{figure}[h!]
    \centering
    \includegraphics{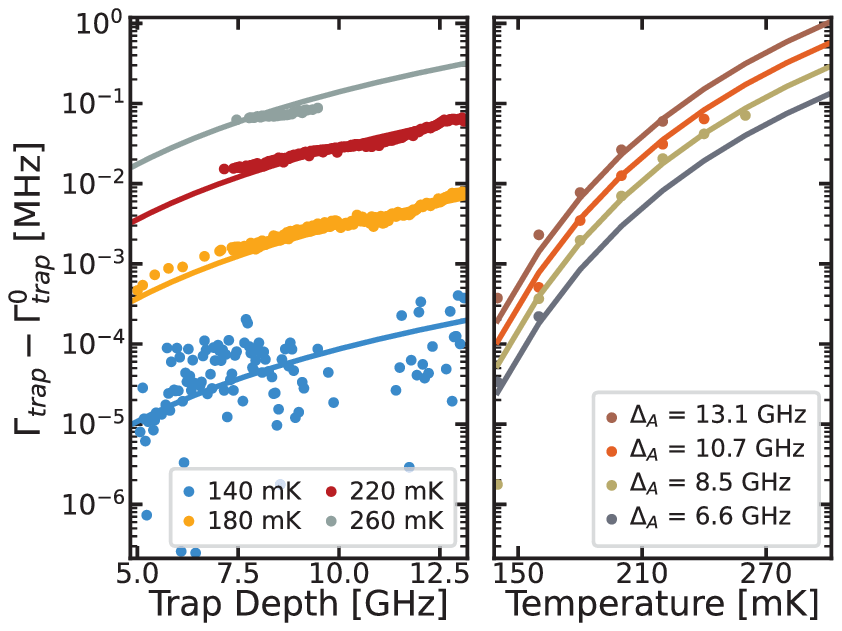}
    \caption{The measured trap rate minus the low temperature trap rate (average of all data less than 80 mK) is shown along with the fit to Eq \ref{eq:subtrap}. We find the scaling factor $\beta = 8.73 \times 10^{15}$ MHz/eV$^3$ and the superconducting gap $\Delta = 185 \mu$eV.}
    \label{fig:subtrap}
\end{figure}
Next, with $\Delta = 185 \mu$eV (from the prior fit) held constant, we fit $\Vert\Gamma_{trap}\Vert$ (Eq. \ref{eq:normtrap}) with the fractional non-equilibrium density $x_{ne}$ as the only free parameter, as shown in Supplementary Figure \ref{fig:normtrap}.
\begin{figure}
    \centering
    \includegraphics{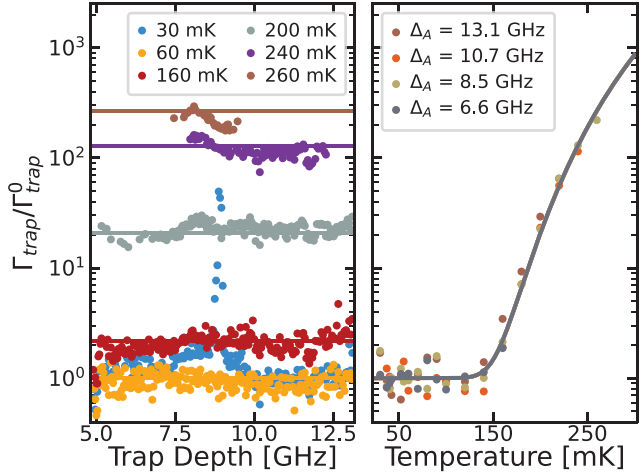}
    \caption{The measured trap rate divided by the low temperature trap rate (average of all data less than 80 mK) is shown along with the fit to Eq \ref{eq:normtrap}. We find the fractional non-equilibrium quasiparticle density $x_{ne} = 8.5\times 10^{-7}$.}
    \label{fig:normtrap}
\end{figure}
The results of this fitting procedure are $\beta = 8.73 \pm 0.68 \times 10^{15}$ MHz/eV$^3$, $x_{ne} = 8.50 \pm 0.10 \times 10^{-7}$, $\Delta = 185.0 \pm 1.5 \mu\text{eV}$ as quoted in the main text.

\section{Derivation of phonon contribution to ABS clearing}
%We start with Eq. \ref{main-eq:rho} of the main text,
As described in Eq. 5 of the main text, we expect the electron-phonon clearing rate to be linear in the phonon density
\begin{equation}
    \Gamma_{phonon}(\Delta_A,T) \propto \rho_{\epsilon \geq \Delta_A}(T).\label{eq:phonon}
\end{equation}
Our first task is to find the density of phonons with energy exceeding the trap depth, $\rho_{\epsilon\geq\Delta_A}(T)$.
Phonons obey Bose-Einstein statistics and we expect them to follow the Debye model.
With $\omega$ as phonon frequency and $\nu$ as the phase velocity, we have the Debye density of states $D(\omega) = \omega^2/2\pi^2\nu^3$.
We obtain the phonon density by integrating $D(\omega)$ over the Bose-Einstein distribution,
\begin{align*}
    \rho_{\epsilon \geq \Delta_A}(T) = &\frac{1}{2\pi^2\nu^3}\int_{\frac{\Delta_A}{\hbar}}^{\omega_D} \frac{\omega^2}{\exp\left(\frac{\hbar\omega}{k_BT}\right) - 1}d\omega\\
    = &\frac{1}{2\pi^2}\left(\frac{k_BT}{\hbar\nu}\right)^3\Bigg[-\left(\frac{\Delta_A}{k_B T}\right)^2
\ln{\left(1-e^{\frac{-\Delta_A}{k_B T}}\right)} + \frac{2\Delta_A}{k_B T} \text{Li}_2\left(e^{\frac{-\Delta_A}{k_B T}}\right)\numberthis\label{eq:big}\\
& \;\;\;\;+ 2\text{Li}_3\left(e^{\frac{-\Delta_A}{k_B T}}\right) + f(\omega_D,T)\Bigg].
\end{align*}
Eq. \ref{eq:big} groups all terms which result from the upper limit of our integration (the Debye frequency $\omega_D$) into one function, 
\begin{equation}
    f(\omega_D,T) = -\left(\frac{\hbar\omega_D}{k_B T}\right)^2
\ln{\left(1-e^{\frac{-\hbar\omega_D}{k_B T}}\right)} + \frac{2\hbar\omega_D}{k_B T} \text{Li}_2\left(e^{\frac{-\hbar\omega_D}{k_B T}}\right) + 2\text{Li}_3\left(e^{\frac{-\hbar\omega_D}{k_B T}}\right).
\end{equation}
There are two materials we may be interested in: silicon and aluminum. Plugging in the appropriate $\omega_D$ for each material ($2\pi\times$ 21.98 THz for Silicon, $2\pi\times$ 15.37 THz for Aluminum), we find that $f(\omega_D,T)$ evaluates to a negligibly small offset in Eq \ref{eq:big} for $T\leq$ 70 K, and near identically zero for our measured temperature range $T\leq300$ mK.
We therefor drop these $\omega_D$ terms.
To relate the phonon density calculated above to ABS clearing, we introduce a coupling factor $C_{ABS\rightarrow bulk}$  which functions as a clearing rate per unit phonon density:
\begin{align*}
    \Gamma_{phonon}(\Delta_A,T) &\equiv C_{ABS\rightarrow bulk}\times\rho_{\epsilon \geq \Delta_A}(\Delta_A,T)\\
    &= \frac{C_{ABS\rightarrow bulk}}{2\pi^2}\left(\frac{k_BT}{\hbar\nu}\right)^3\Bigg[-\left(\frac{\Delta_A}{k_B T}\right)^2
\ln{\left(1-e^{\frac{-\Delta_A}{k_B T}}\right)} + \frac{2\Delta_A}{k_B T} \text{Li}_2\left(e^{\frac{-\Delta_A}{k_B T}}\right)\numberthis\label{eq:full}\\
& \;\;\;\;+ 2\text{Li}_3\left(e^{\frac{-\Delta_A}{k_B T}}\right) + f(\omega_D,T)\Bigg].
\end{align*}
The model presented in Eq. 6 of the main text combines all the preceeding constants into a single parameter
\begin{equation}
    \alpha = \frac{k_B^3}{2\pi^2\hbar^3\nu^3}C_{ABS\rightarrow bulk}.
\end{equation}
%Using the speed of sound for bulk aluminum at room temperature and $\alpha = 38.51$ MHz/K$^3$ from the main text, we obtain $C_{ABS\rightarrow bulk} =  89.6 $ kHz/$\mu$m$^{-3}$, which we interpret as the rate of clearing quasiparticles from ABS to the bulk per unit phonon density.